# TWO AUTOMATED TECHNIQUES FOR ANALYZING AND DEBUGING MPI-BASED PROGRAMS

Sultan Aljahdali
Computer Science Department
College of Computers and Information Technology
Taif, Saudi Arabia
aljahdali@tu.edu.sa

Mosaid Al Sadhan
King Saud University
Riyadh, Saudi Arabia
sadhan3@gmail.com

Alaa Ismail Elnashar
Computer Science Department
College of Computers and Information Technology
Taif, Saudi Arabia
a.ismail@tu.edu.sa

**Abstract**

*Message Passing Interface (MPI) is the most commonly used paradigm in writing parallel programs since it can be employed not only within a single processing node but also across several connected ones. Data flow analysis concepts, techniques and tools are needed to understand and analyze MPI-based programs to detect bugs arise in these programs. In this paper we propose two automated techniques to analyze and debug MPI-based programs source codes.*

## 1 INTRODUCTION

Three main models for programming parallel architectures are currently used. These models are message-passing paradigm (MPI) [25], shared memory programming model, and Partitioned Global Address Space (PGAS) programming model [7]. Message Passing Interface (MPI) is the most commonly used paradigm in writing parallel programs since it can be employed not only within a single processing node but also across several connected ones. MPI standard has been designed to enhance portability in parallel applications, as well as to bridge the gap between the performance offered by a parallel architecture and the actual performance delivered to the application [8]. It offers several functions such as point-to-point rendezvous-type send/receive operations, logical process topology, data exchange, gathering and reduction operations and synchronization.

Shared memory programming model allows a simpler programming of parallel applications, as the control of the data location is not required. OpenMP [9] is the most suitable solution used for shared memory programming, as it allows an easy development of parallel applications through compiler directives. Hybrid systems, with both shared/distributed memory, such as multi-core clusters, can be programmed using MPI combined with OpenMP. However, this hybrid model can make the parallelization more difficult and the performance gains could not compensate for the effort [7].

Partitioned Global Address Space (PGAS) programming model combines the main features of the message passing and the shared memory programming models. In PGAS model, each process has its own private memory space, as well as an associated shared memory region of the global address space that can be accessed by other processes. It also allows shared memory-like programming on distributed memory systems. Moreover, as in MPI, PGAS allows the exploitation of data locality as the shared memory is partitioned among the processes in regions, each one with affinity to the corresponding process. Several implementations such as Parallel Virtual Machine (PVM) [23] and MPICH2 [24] are now available and can be used in writing MPI-based programs. Parallel Virtual Machine (PVM), is a software package that permits a heterogeneous collection of UNIX or Windows computers hooked together by a network to be used as a single large parallel computer. MPICH2 is a high performance and widely portable implementation of MPI standard. It efficiently supports different computation and communication platforms. In this paper we focus on using MPICH2 for Windows platforms. Most of parallel Application programmers focus only on the constructive part of creating a parallel algorithm for a particular problem and how to implement it, but ignore the issues of debugging [2]. Parallel programming adds new types of bugs caused by the interaction of multiple communicated parallel processes. These parallel bugs 'Heisen bugs' [14] are difficult to be detected and resolved due to the nondeterministic nature of the running parallel processes which makes the bugs may disappear when one attempt to detect them.

In this paper we present the implementation of two automated techniques that assist in the analysis of MPI-based programs and detecting of some parallel programming bugs. The paper is organized as follows: section 2 gives a brief idea about the related work. In section 3, we discuss the analysis of MPI-based programs. Section 4 presents the automated analysis technique. In section 5, the automated debugging technique is presented.

## 2 RELATED WORK

Many techniques are used for locating parallel programming bugs. The most commonly used techniques,

are dynamic analysis, static analysis and model-based test. Dynamic analysis implies the necessity of launching an application and executing different sequences of operations to analyze the program behavior. As an example, Intel Message Checker (IMC) [13] performs a post-mortem analysis by collecting all information on MPI calls in a trace file. After program execution, this trace file is analyzed by a separate tool or compared with the results from previous runs. The traditional serial debuggers can also be used to debug MPI-based applications by setting breakpoints to investigate a specific state. The Debugger allows the programmer to single-step through his running application to test a process against a specific fault. GNU debugger (gdb) [21] and Data Display Debugger (ddd) [1] can be used to debug MPI-based parallel applications. MPI-CHECK [11], Umpire [15] and MARMOT [3] debuggers are effective in detecting some types of software bugs at runtime but still poor to detect semantics-related bugs [19].

Static analysis approach handles only the source code without its execution. This approach can be useful to determine detailed and full coverage of the analyzed code. In case of MPI-based programs, static analysis can detect errors that may not appear during real program execution, and hence it can complement dynamic analysis to discover more bugs. It requires an intermediate source code representation such control flow graphs CFG [17] as in case of data flow testing. This means that extra effort has to be done in building CFG representing MPI-based programs MPI-CFG [6] since the ordinary CFG does not demonstrate most of MPI constructs like inter-process communication and synchronization edges.

Model-based testing is a testing approach in which test cases are derived from a model that describes the system under test. Practically, model-based test works only for small base blocks of an application. In most cases it is very difficult to automatically build a model on the basis of the code; the manual creation of models is a hard and error prone process. It also suffers from the problem of quick extension of state space. For MPI-based programs, this approach would require that programmers build, either manually or automatically, a model of their applications in a language such as MPI-SPIN [26], Zing [20] and PPL[18].

## 3 ANALYSIS OF MPI-BASED PROGRAMS

MPI-based programs are coded in a special manner, in which each process executes the same program with unique data. All parallelism is explicit; the programmer is responsible for identifying parallelism and implementing parallel algorithms using MPI constructs.

### 3.1 MPI Programming Model

MPI is available as an open sources implementations on a wide range of parallel platform. In MPICH2 implementation the MPI-based source program is compiled and linked with the MPI libraries to obtain the executable. The user issues a directive to the operating system that places a copy of the executable program on each processor, the number of processes is provided within the user directive. Each processor begins execution of its copy of executable. Each process can execute different statements by branching within the program based on a unique rank "process identifier". This form of MIMD programming is frequently called Single-program multiple-data SPMD. Each process has its own local memory address space; there are no shared global variables among processes. All communications are performed through special calls to MPI message passing routines. MPI uses objects [13] called communicators and groups to define which collection of processes may communicate with each other. A communicator must be specified as an argument for most MPI routines. An MPI-based program consists of four parts. The first one is the MPI include file which is required for all programs/routines which make MPI library calls. The second part is responsible for initializing MPI environment. Special function calls are used for initializing and terminating the MPI environment. The function MPI_Init initializes the MPI execution environment. This function must be called only once in an MPI-based program before any other MPI functions. MPI_Comm_size, determines the number of processes in the group associated with a communicator. Generally used within the communicator MPI_COMM_WORLD to determine the number of processes being used by the application. MPI_Comm_rank, determines the rank of the calling process within the communicator. The third part is the body of program instructions, calculations, and message passing calls. The last one is terminating MPI environment by calling the function MPI_Finalize. MPI provides several routines used to manage the inter-process communications via send / receive operations.

### 3.2 Data Flow of MPI-based Programs

Figure 1 shows a pseudo code of an MPI-based program. Running the executable of the listed code several times using three process may yields one of two outputs, one of them indicates that the value 4 is sent from process 1 to process 0 and the sum value is 7, the other one indicates that the value 14 is sent from process 2 to process 0 and the sum value is 17. The order of these outputs is unpredictable. This situation reflects the non-deterministic behavior of program execution.

These results demonstrate that the affected statements are not the only affected ones but also there are some other statements that should be encountered, it can be noticed that this analysis fails to detect the effect of the definition of "x" on the computation of "sum" in line 6, as shown in table 2.

Variable definitions like " sum = 3 " in line 1, are shared in SPMD programs without a communicator, so they are considered as global variables. On the other hand, variables defined within each process section, like "sum=sum + received" in line 6 can't be shared outside this section unless an appropriate communicator is used. These variables are considered as local variables.

```
1. sum=3
2. Initialize MPI environment.
2. Determine the number of MPI processes
   and their identities.
3. if  myid=0 then
4.    Receive "received" from any process
5.    Receive "sender_id" from any process
6.    sum = sum + received
7.    x0 = sum
8. endif
9. if myid=1 then
10.   x = 5
11.    if x<0  then
12.        x = x +1
13.    else
14.        x = x-1
15.    end if
16.   Send "x" to process 0
17.   process_id = myid
18.   Send "process_id" to process 0
19. endif
20. if  myid = 2  then
21.    x = 7
22.    x = x * 2
23.    Send "x" to process 0
24.    process_id = myid
25.    Send "process_id" to process 0
26. endif
27. Finalize MPI environment.
28. END
```

Figure 1. pseudo MPI –based code

Table1

| Case | Affected statements |
|---|---|
| definition of sum in line 1 | 6 sum=sum + received |
| definition of sum in line 6 | 7 x0 = sum |
| definition of x in line 10 | 11 if x<0 then<br>12 x = x +1<br>14 x = x – 1 |
| definition of x in line 12 | 16 Send "x" to process 0 |
| definition of x in line 14 | 16 Send "x" to process 0 |
| definition of x in line 21 | 22 x = x * 2 |
| definition of x in line 22 | 25 Send "x" to process 0 |

Table 2

| Case | Statements should be encountered |
|---|---|
| definition of x in lines 12, 14 | 4 Receive "received".. |
| definition of x in line 22 | |

## 4 AUTOMATED ANALYSIS TECHNIQUE

Static data flow analysis is a technique for gathering information about the possible set of values calculated at various points in a sequential program. Data flow analysis techniques represent a program by its control flow graph, CFG, which consists of a set of nodes and edges. Each node represents a basic block which is a set of consecutive statements of the program, and each edge represents the control flow between these blocks. The goal of analysis techniques is to identify which definitions of program variables can affect which uses. To build CFG, the analyzed program is divided into a set of basic blocks, the set of edges connecting these blocks according to the flow of control is generated. The constructed CFG is then used by the static analyzer to identify the def-use associations among the blocks. CFG is used to determine those parts of a program to which a particular value assigned to a variable might propagate. This can be done by generating two sets, $dcu(i)$ and $dpu(i,j)$ [17] for program variables. These two sets are necessary to determine the definitions of every variable in the program and the uses that might be affected by these definitions. The set $dcu(i)$ is the set of all variable definitions for which there are def-clear paths to their c-uses at node $i$.

$dpu(i,j)$ is the set of all variable definitions for which there are def-clear paths to their p-uses at edge $(i,j)$ [16]. Using information concerning the location of variable definitions and references, together with the "basic static reach algorithm" [10], the two sets can be determined. The basic static reach algorithm is used to determine the sets reach($i$) and avail($i$). The set reach($i$) is the set of all variable definitions that reach node $i$. The set avail($i$) is the set of all available variables at node $i$. This set is the union of the set of global definitions at node $i$ together with the set of all definitions that reach this node and are preserved through it. Using these two sets, the sets $dcu(i)$ and $dpu(i,j)$ are constructed from the formula:

$$dcu(i) = reach(i) \cap c-use(i)$$
$$dpu(i,j) = avail(i) \cap p-use(i,j)$$

This technique fails to demonstrate a correct analysis for MPI-based programs. The SPMD nature needs to be modeled correctly in the program representation to be considered during static program analysis; this can be achieved by using a special data structure that can represent sequential flow, parallel flow and synchronization in explicitly MPI-based programs.

### 4.1 MPI-CFG Construction Challenges

Building a CFG representing MPI-based programs (MPI-CFG) is restricted by the following challenges:

1. Processes in MPI-based programs are declared by using the ordinary conditional "IF" statement depending on the process identifier. This will make confusion during dividing the program into basic blocks, and also during the process of generating edges which will badly affect the operations of static analyzer. "IF" statements that are used to declare processes must be treated in a special manner rather than that is used in treating "IF" statements used within the body of each process as shown in figure 1, line 9 and line 11.
2. MPI-based program is executed by more than one process, each process has its local memory address space; there is no shared global variables among these processes except that are defined before the initialization of the MPI execution environment. This requires identifying both local and global variables.
3. Def-use association among process can be achieved only by calling inter-process communication message passing routines (send/ receive). This implies constructing extra edges that represent these constructs.

**4.2      Implementation of MPI-CFG Construction**

Now we present our technique to build the MPI-CFG. This flow graph resembles the synchronized flow graph [5], program execution graph [22], parallel flow graph [12], and parallel program flow graph PPFG [4]. The technique works as follows:

1. MPI-based program statements identification.
  In this phase, each program statement is assigned a unique number "type" to be identified from the other statements of the program. The phase must check for the following:
  a) If the statement "Call MPI_Comm_rank( XX,YY,ZZ) " is encountered, it is assigned its type and the second parameter YY which indicates the variable name that will be used to identify the parallel processes is recorded as "special_id".
  b) The assigned type of conditional "IF" statements depends on the recorded "special_id"; if the value of "special_id" appears in the condition, this means that the encountered "IF" statement is used to declare a process, otherwise, it is an ordinary conditional statement.
  The output of this phase is an intermediate representation of the source code. It contains numbered statements of the input program associated with their types and the recorded "special_id".
2. Building Basic Blocks
  This phase uses the output of the previous phase to build the program basic blocks. We construct two extra special types of basic blocks called "message block" and "finalize block". A message block is either "receive block" or "send block". A basic block that has at most one communication statement at its start is said to be "receive block". This block is constructed if the statement call MPI_Recv( ) is encountered. The "send block" has at most one communication statement at its end. This block is constructed if the statement call MPI_Send( ) is encountered. The "finalize block" is constructed if call MPI_Finalize( ) statement is encountered. During building basic blocks the program variables, their block numbers and their status (def, c-use, or p-use) are also recorded.
  At the termination of this phase another version of the input MPI-based program is generated. This version contains the statement and block number for each program statement. All the required information about the variable names and the parameters of send/receive constructs are also recorded.
3. Generating Edges.
  This phase connects the basic blocks generated in the previous phase with the appropriate edges. The edges are classified into three categories, sequential, parallel, and synchronization edges. Sequential edges indicate a possible flow from a block to another one. This type of edges is used to connect the basic blocks within each process as the ordinary sequential flow edges. Parallel edges represent the parallel control flow at both process declaration and termination points. Synchronization edges represent the inter-process communication via send/receive operations. Synchronization edges are generated by matching the parameters of call MPI_Recv( ) and call MPI_Send( ) recorded in the second phase. The output of this phase is the MPI-CFG. Figure 2 shows the MPI-CFG of the real code corresponds to the pseudo code listed in Figure 1.

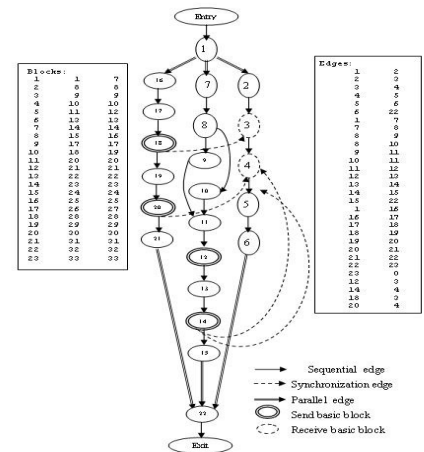

Figure 2. MPI Control flow graph

We applied the ordinary data flow analysis technique described above on the constructed MPI-CFG with some modifications to handle the nature of MPI-based programs.

**5      Automated Debugging Technique**

There are four possible types of anomalies that may occur due to the inter-process communication of parallel program processes.

The first type of anomalies occurs when a process waits for itself.

The second type is the deadlock, which occurs if a process is suspended waiting for a value(s) to be sent from another one that is either not exists or waits for the first one.

The third one is the run-time non-determinacy that occurs if one WAIT corresponds to more than one SEND. This anomaly can be noticed only at run time. In this type of errors the program returns different results in repeated executions for the same input data due to the non-deterministic nature of parallel programs.

The last type of anomalies arises if there is a send primitive without its corresponding wait primitive.

In this section we present a new technique to detect these anomalies. The technique accepts a program that consists of a set of parallel processes with the synchronization primitives as its input, and then traces the input program to detect any inter-process communication anomalies, if any. The steps of the technique are described below:

1. Input the given source program.
2. Isolate Wait(s) and Send(s). In this step, two files are created; one records the wait statements and the other records the send statements within the input program. Each file contains, for every wait/send, the number of the process which includes the wait/send statement, the statement number, variable name that this process wait for/send to another process, and the number of the other process that will send/receive that variable.
3. Scan waits and sends files in the following manner:
    a) For each wait primitive in the waits file, if the number of the process which includes the wait statement is the same as the number of the other process that will send a certain variable then an anomaly of the first class "a process waits for itself" is detected.
    b) The detection of deadlocks can be achieved by finding those wait primitives that have no corresponding send ones in the sends file. In this case the line number of this statement and also the process number is recorded.
    c) Each send primitive is compared with all waits to determine whether there is a wait/send matching or not. If there is no matching, the line number of these statements and also the processes numbers are recorded.
    d) Collecting the recorded information generated from the previous steps, the anomalies report is generated.

The flow chart of this technique is shown in figure 3.

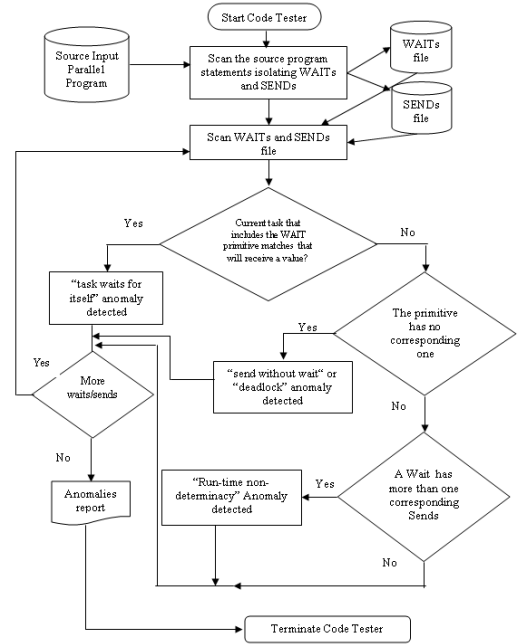

Figure 3. Automated debugging process

## 6    CONCLUSION

Unlike sequential programs, data flow analysis of MPI-based programs requires extra effort. Applying existing concepts, techniques and tools used to analyze sequential programs on MPI-based programs fails to report a correct program analysis. These techniques require some modifications to handle the SPMD nature of MPI programs. We have implemented a technique to extend the program CFG to represent the MPI-based programs MPI-CFG. The static analyzer uses the constructed graph to generate the program analysis report. We have implemented the techniques of building message basic blocks, constructing parallel edges, and also constructing synchronization edges represent send/receive constructs to produce a correct MPI-based program representation. Also, a technique for detecting parallel bugs is proposed.

In future, we hope to implement the construction of synchronization edges for all MPI inter-process communication constructs.